%% file: main.tex
\documentclass[sigconf]{acmart}
\makeatletter
\def\@ACM@checkaffil{
    \if@ACM@instpresent\else
    \ClassWarningNoLine{\@classname}{No institution present for an affiliation}%
    \fi
    \if@ACM@citypresent\else
    \ClassWarningNoLine{\@classname}{No city present for an affiliation}%
    \fi
    \if@ACM@countrypresent\else
        \ClassWarningNoLine{\@classname}{No country present for an affiliation}%
    \fi
}
\makeatother

\usepackage{multirow}
\usepackage{xspace}
\usepackage{enumitem}
\usepackage{graphicx}
\usepackage{subfigure}

\newcommand{\paratitle}[1]{\noindent\textbf{#1}}
\newcommand{\trans}{{\mkern-1.5mu\mathsf{T}}}
\newcommand{\model}{GraphPro\xspace}
\AtBeginDocument{%
  \providecommand\BibTeX{{%
    \normalfont B\kern-0.5em{\scshape i\kern-0.25em b}\kern-0.8em\TeX}}}

\setcopyright{acmcopyright}

\begin{document}

\title{GraphPro: Graph Pre-training and Prompt Learning \\ for Recommendation}



\author{Yuhao Yang}
\affiliation{University of Hong Kong}
\email{yuhao-yang@outlook.com}

\author{Lianghao Xia}
\affiliation{University of Hong Kong}
\email{aka\_xia@foxmail.com}

\author{Da Luo}
\affiliation{Wechat, Tencent}
\email{lodaluo@tencent.com}

\author{Kangyi Lin}
\affiliation{Wechat, Tencent}
\email{plancklin@tencent.com}

\author{Chao Huang}
\authornote{Chao Huang is the corresponding author.}
\affiliation{University of Hong Kong}
\email{chaohuang75@gmail.com}

\renewcommand{\shortauthors}{Yuhao Yang et al.}

\begin{abstract}
GNN-based recommendation systems have excelled at capturing complex user-item interactions through multi-hop message passing. Nevertheless, these methods often fail to account for the dynamic nature of user-item interactions, leading to challenges in adapting to changes in user preferences and the distribution of new data. Consequently, their scalability and performance in real-world dynamic settings are constrained. In our study, we introduce \model, a framework that merges dynamic graph pre-training with prompt learning in a parameter-efficient manner. This innovative blend enables GNNs to adeptly grasp both enduring user preferences and transient behavior changes, thereby providing precise and up-to-date recommendations. \model tackles the issue of changing user preferences through the integration of a temporal prompt mechanism and a graph-structural prompt learning mechanism into the pre-trained GNN architecture. The temporal prompt mechanism imprints time-related information onto user-item interactions, equipping the model to inherently assimilate temporal dynamics, while the graph-structural prompt learning mechanism allows for the application of pre-trained insights to new behavior dynamics without continuous retraining. We also introduce a dynamic evaluation framework for recommendations to better reflect real-world situations and narrow the offline-online discrepancy. Our comprehensive experiments, including deployment in a large-scale industrial context, demonstrate the effortless plug-in scalability of \model alongside various leading recommenders, underscoring the superiority of \model in effectiveness, robustness, and efficiency. The implementation details and source code of our \model are available in the repository at \color{blue}\url{https://github.com/HKUDS/GraphPro}.
\end{abstract}

\maketitle

\input{intro}
\input{method}

\input{exp}
\input{relate}

\section{Conclusion}
This study introduces \model, a new framework combining dynamic graph pre-training with prompt learning, aiming to substantially enhance the adaptability and scalability of recommender systems that are sensitive to temporal dynamics. \model uses a temporal prompt mechanism to transfer knowledge from pre-trained models to recommendation tasks on fresh data. Its graph-structured prompt learning features an adaptive gating mechanism that includes vital contextual information, easing fine-tuning and responsiveness to behavioral shifts. Through an extensive array of experiments conducted on a range of real-world datasets, we have empirically validated that \model consistently outpaces contemporary state-of-the-art baselines, delivering exceptionally accurate dynamic recommendations across various temporal contexts. Looking ahead, our research will delve into the interpretability aspect of \model, particularly focusing on the prompt graph edges.

\clearpage
\bibliographystyle{ACM-Reference-Format}
\bibliography{references_backup}

\clearpage
\input{appendix}
\end{document}

%% file: intro.tex
\section{Introduction}

Recommender systems are integral to numerous Web platforms, assisting users in navigating through the overwhelming amount of information by suggesting relevant items. In recent years, graph neural networks (GNNs) have emerged as powerful tools for modeling user-item interactions in recommendation tasks, enabling effective representation learning on graph-structured data. By treating users and items as nodes and their interactions as edges, GNNs can capture intricate multi-hop relationships between users and items, facilitating the generation of personalized recommendation results.

Previous studies in GNN-augmented recommendation have centered on crafting effective message passing strategies to delineate the collaborative relations between users and items~\cite{gcmc,pinsage,ngcf}. These efforts seek to harness GCN's potential to unravel high-order connections within the user-item interaction graph. Follow-up investigations have delved into streamlining the message-passing approach~\cite{chen2020revisiting,lightgcn}, diminishing GNN model intricacies~\cite{ultragcn,shen2021powerful}, and advancing sampling method quality~\cite{huang2021mixgcf}. Lately, the field has progressed with the integration of self-supervised learning (SSL) into GNN frameworks for recommendation~\cite{ren2023sslrec}. These cutting-edge methodologies~\cite{sgl,cai2023lightgcl,simgcl} predominantly utilize the InfoNCE loss function~\cite{infonce} to align contrastive perspectives, thus enhancing the performance of the LightGCN architecture~\cite{lightgcn}.

\begin{figure}[t]
    \centering
    \includegraphics[width=0.93\linewidth]{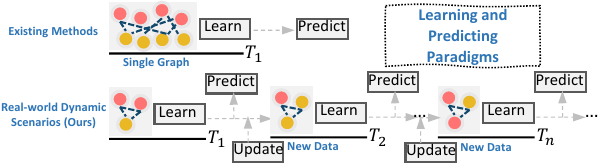}
    \vspace{-0.15in}
    \caption{Our dynamic recommendation setting compared to the vanilla single-graph training in existing methods.}
    \label{fig:intro}
    \vspace{-0.2in}
\end{figure}

Although these methods have demonstrated impressive performance, they have primarily focused on static scenarios (Figure~\ref{fig:intro} upper), disregarding the dynamic nature of recommendation. In real-world scenarios (Figure~\ref{fig:intro} lower), the recommendation system operates in a dynamic setting where the model continuously learns from newly arriving data and predicts for the current time~\cite{xiao2023reconsidering,yang2023generic}. However, existing methods are mainly designed for single-graph training and evaluation, leading to a degradation in recommendation dynamics and widening the offline-online gap~\cite{ji2023critical}. Furthermore, the arrival of new data may exhibit distribution shifts, further complicating the task of making accurate recommendations for graph-based recommendation models without incorporating useful contextual information for the newly arrived data. These challenges significantly limit the scalability of existing models and hinder their ability to adapt to evolving user preferences in a timely manner. This is critical for providing up-to-date and precise recommendations in dynamic environments.


To address these challenges, we present a simple and effective framework called \model, which combines parameter-efficient and dynamic \underline{Graph} Pre-Training with \underline{Pro}mpt Learning for recommendation. Our method begins with pre-training Graph Neural Networks (GNNs) on extensive historical interaction data, enabling them to capture long-term user preferences and item relevance. This pre-training phase assimilates knowledge from a substantial amount of historical interactions. Subsequently, during the fine-tuning phase on more recent target data, our model swiftly adapts to evolving user preferences and captures short-term behavior dynamics. This is achieved through a prompt learning schema, which facilitates effective knowledge transfer.



To ensure the effective handling of evolving user preferences by the pre-trained GNN, our \model\ framework seamlessly integrates a temporal prompt mechanism and a graph-structural prompt learning mechanism. This integration allows for the injection of time-aware context from new data, enabling the model to adapt to changing user preferences. Inspired by advancements in relative positional encoding techniques~\cite{alibi,rope}, we meticulously design a dedicated temporal prompt mechanism that aligns with the message aggregation layer of GNNs. Within this prompt mechanism, we encode time information on interaction edges as part of the normalization term for aggregation, all in a parameter-free manner. This innate capability allows the model to naturally incorporate temporal information without the need for additional fine-tuning. By incorporating temporal awareness into the pre-trained graph model, we empower the model to effectively capture vital signals that are highly relevant to the evolving user preferences.


Moreover, our graph-structural prompt learning mechanism facilitates the seamless transfer of knowledge from the pre-trained model to downstream recent recommendation tasks. This framework eliminates the need for continuous incremental learning of the pre-trained model, enabling the transfer of pre-trained knowledge to any future time period to effectively adapt to behavior dynamics. In this mechanism, we include the newly generated interaction edges between the fine-tuning time and the pre-training time as prompt edges, providing the pre-trained model with essential contextual information for the fine-tuning process. Rather than undergoing extensive training, we perform a single non-training forward pass on the prompt edges, prompting the pre-trained model to adapt to the distribution shift of node representations and effectively adjust its predictions. It's worth mentioning that our \model\ is model-agnostic and parameter-efficient, making it easy to integrate into existing GNN recommenders as a plug-in enhancement. In summary, the main contributions of our work are as follows:

\begin{itemize}[leftmargin=*]

\item We emphasize the importance of effectively and scalably pre-training and fine-tuning graph-based recommenders to accommodate time-evolving user preferences, thus facilitating up-to-date and accurate recommendations in dynamic environments.

\item We introduce \model, which effectively handles evolving user preferences through the pre-training and fine-tuning of GNNs. The proposed prompt learning paradigm enables the transfer of relevant knowledge from the pre-trained model to downstream recommendation tasks in both temporal and structural ways.

\item Furthermore, we introduce a snapshot-based dynamic setting for recommender system evaluation, which offers a more realistic approximation to real-world recommendation scenarios compared to the traditional single-time testing approach.


\item We conduct experiments on diverse datasets to showcase the robustness, efficiency, and performance advantages of \model. To further validate the effectiveness of our framework, we also present an industry deployment on a large-scale online platform.


\end{itemize}


%% file: method.tex
\section{PRELIMINARIES}

We define the task of pre-training and fine-tuning GNNs for recommendation. We denote the user set as $\mathcal{U}$ and the item set as $\mathcal{I}$. In the context of collaborative filtering, a typical graph structure, constructed using existing methods~\cite{lightgcn}, can be represented as $\mathcal{G}=(\mathcal{V},\mathcal{E})$, where $\mathcal{V}=\mathcal{U}\cup\mathcal{I}$ represents the set of all nodes in user-item interaction graph $\mathcal{G}$. The edges in $\mathcal{E}$ correspond to interactions between users and items, with a value of $y_{u,i}=1$. \\\vspace{-0.12in}

In order to provide recommendations at time slot $T_1$, we gather historical user-item interactions to construct a graph $\mathcal{G}_{1} = (\mathcal{V}, \mathcal{E}_{1})$, where $\mathcal{E}_{1}$ represents the user-item interactions collected before $T_1$. Existing stationary graph collaborative filtering models typically train the model from scratch using the complete dataset $\mathcal{G}_{1}$. The objective is to optimize time-specific model parameters $\Theta_{1}$ by maximizing the likelihood of generating accurate recommendations:
\begin{equation}
\underset{\Theta_{1}}{\text{argmax}}\, P_{f_{\Theta_{1}}}(y_1| \mathcal{G}_{1}).
\end{equation}

\paratitle{Dynamic Learning in Recommender Systems}. In real-world applications, the evaluation of recommenders goes beyond the simplistic static setting commonly used in existing collaborative filtering method~\cite{ngcf,lightgcn}. In practice, researchers assess the long-term performance of models by deploying them in a live-update environment, as discussed in~\cite{you2022roland}, where new user-item interactions are continuously generated over time. The model is specifically designed to make ongoing predictions for future user-item interactions based on this evolving data. Formally, the model should have initial weights $\Theta_{n-1}$ corresponding to different time intervals $T_n$, which are then updated through learning on new interactions $\mathcal{G}_{n}$ to enhance the accuracy of up-to-date predictions.
\begin{equation}
\underset{\Theta_{n}}{\text{argmax}}\, P_{f_{\Theta_{n}}}(y_{n}| \mathcal{G}_{n};\Theta_{n-1}).
\end{equation}

In this study, we draw inspiration from the work~\cite{pareja2020evolvegcn} and employ a series of graph snapshots to simulate practical dynamic recommendation scenarios. These graph snapshots are represented as $[\mathcal{G}_1,\mathcal{G}_2,...,\mathcal{G}_N]$, where we have a total of $N$ snapshots corresponding to different time intervals. Each graph snapshot consists of subsets of users and items from a global set, and the interactions between them evolve over time. Specifically, $\mathcal{G}_n = (\mathcal{V}_n\subset\mathcal{V}, \mathcal{E}_n)$, where $\mathcal{V}_n$ represents the nodes in the snapshot and $\mathcal{E}_n$ denotes the time slot-specific interaction edges. It is important to emphasize that snapshots $\mathcal{G}_n$ are collected within the time period between two consecutive snapshots, namely $[T_{n-1},T_n]$.

\section{Methodology}
\begin{figure}
    \centering
    \includegraphics{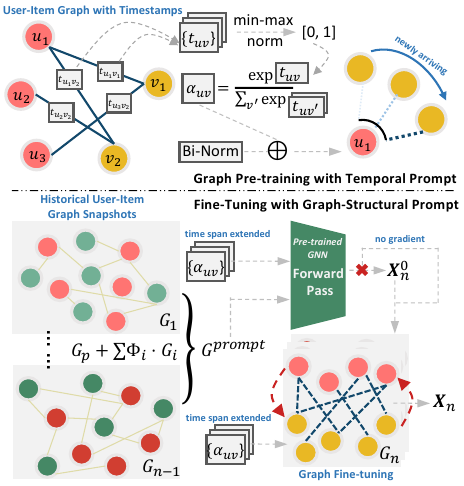}
    \vspace{-0.15in}
    \caption{Overall framework of \model.}
    \label{fig:fra}
    \vspace{-0.2in}
\end{figure}

In this section, we provide the technical details of our proposed \model\ framework, depicted in Figure~\ref{fig:fra}, which illustrates its architecture. We introduce two crucial components: graph pre-training with a temporal prompt mechanism and a graph-structural prompt-enhanced fine-tuning mechanism. These components are specifically designed to enhance the performance and scalability of GNN-based recommenders in dynamic recommendation scenarios.

\subsection{Graph Pre-training with Temporal \\ Prompt Mechanism}
In practical recommendation scenarios, user-item interaction data continues to accumulate over time. Online platforms like Amazon and TikTok constantly receive new user purchases and video watch logs, respectively, on a daily basis. In such dynamic environments, the availability of fresh user-item interactions provides valuable information that can be leveraged to guide pre-trained models in adapting to time-evolving user preferences and providing continuously up-to-date recommendations in dynamic settings.

\subsubsection{\bf Temporal Prompt Mechanism}
In our framework, we introduce a temporal prompt mechanism to incorporate time-aware contextual information from the latest user preferences and behaviors. This mechanism allows for personalized and timely recommendations by considering the temporal dynamics of user-item interactions. To capture the temporal sequence of user-item interactions, we propose a relative time encoding scheme, which enables us to incorporate temporal information into graph convolutions. By encoding the relative time between interactions, the model can explicitly capture the temporal dependencies and changes in user preferences that are reflected in the newly arrived data.

Our temporal prompt mechanism offers two significant advantages over existing time encoding techniques when it comes to capturing user behavior dependencies across different time slots.

\begin{itemize}[leftmargin=*]

\item \textbf{Generalization}. Unlike the use of absolute positional embeddings in models like BERT~\cite{devlin2018bert}, our mechanism takes inspiration from recent advancements in sequence modeling in NLP and leverages a relative positional encoding. Absolute positional embeddings have limited generalization capabilities across continuous time steps, which is problematic in our dynamic recommendation setting. These embeddings are trained on sequences with varying lengths but struggle to handle sequences beyond the trained lengths during fine-tuning and prediction. This leads to a distribution gap between the pretraining phase, where the model learns from fixed-length data, and the fine-tuning and prediction phase, where longer future time steps are encountered. \\\vspace{-0.12in}

\item \textbf{Scalability}. Our temporal prompt design avoids the need to add a fixed-length positional embedding to node representations. Instead, it generates relative temporal-aware weights that can be seamlessly integrated with message passing. This design allows our pretrained GNN to be easily applied to longer-range graph structures during fine-tuning and testing, greatly improving scalability for dynamic recommendation tasks.

\end{itemize}

\subsubsection{\bf Temporal Prompt-enhanced Graph Convolutions}
In order to effectively capture the temporal dynamics of user-item interactions in our model, we have implemented a temporal prompt and incorporated relative time encoding into our graph convolutions. This enables our model to consider the most recent contextual signals from the new data, and adapt to evolving user preferences over time. Within the context of our user-item interaction graph $\mathcal{G}$, the edge attributes consist of Unix timestamps denoted as $\boldsymbol{t}^{\text{unix}}$. These timestamps represent the exact moments when users $u$ interacted with items $v$. To prepare these timestamps for encoding in our model, we convert them into relative time steps by dividing them by a fixed time interval $\tau$. This time interval, which is a hyperparameter, can be defined with a resolution of either hour, day, or week. As a result, for any given edge $e_{u,v}$ in the graph, its corresponding timestep attribute can be computed as follows:
\begin{equation}
\label{eq:div}
   t_{u,v} = f_{\text{div}}(t^{\text{unix}}_{u,v}, \boldsymbol{t}^{\text{unix}}) = \lfloor \frac{t^{\text{unix}}_{u,v} - \min(\boldsymbol{t}^{\text{unix}})}{\tau} \rfloor,
\end{equation}
Here, $t^{\text{unix}}_{u,v}$ is the Unix timestamp assigned to the edge $e_{u,v}$, and the $\lfloor*\rfloor$ notation denotes the floor operation. To avoid the influence of specific numerical scales and ensure uniformity, we normalize these time attributes $\boldsymbol{t} = {t_{u,v}|e_{u,v}=1}$ to the range of [0, 1].
\begin{equation}
\label{eq:norm}
    \boldsymbol{t} = \frac{{\boldsymbol{t} - \min(\boldsymbol{t})}}{{\max(\boldsymbol{t}) - \min(\boldsymbol{t})}}.
\end{equation}
To consider the temporal information among the neighbors during message aggregation in our GNNs, we apply the softmax function to the time attributes $t_{u,v}$ of the first-order neighbors on the graph.
\begin{equation}
\label{eq:tsoftmax}
    \alpha_{u,v} = \frac{{e^{t_{u,v}}}}{{\sum_{v^\prime \in \mathcal{N}_u} e^{t_{u,v^\prime}}}},
\end{equation}
To enable dynamic time-aware graph neural network (GNN) for recommendation pretraining, we introduce an additional normalization term, $\alpha_{u,v}$, into the message passing step of LightGCN. In this case, $\mathcal{N}_u$ represents the neighbors of node $u$, and $\alpha_{u,v}$ encodes the weight for aggregating information from node $v$ to node $u$.
\begin{equation}
\label{eq:agg}
\mathbf{x}_u^{(l)} = \sum_{v\in\mathcal{N}_u}(\frac{1}{2\sqrt{|\mathcal{N}_u||\mathcal{N}_v|}}+\frac{\alpha_{u,v}}{2})\mathbf{x}_v^{(l-1)},
\end{equation}
We introduce a normalization term and apply mean-pooling to incorporate time-aware normalization into the original bidirectional graph normalization while preserving embedding magnitude. \\\vspace{-0.12in}

\noindent \textbf{Adaptability and Efficiency}. The incorporation of the time-aware normalization term $\alpha_{u,v}$ into the message passing of LightGCN enhances the GNN's adaptability to evolving user-item interactions over time. By giving more weight to interactions that are closer in time during neighbor aggregation, the model becomes more attentive to the dynamic nature of user-item interactions and assigns higher importance to recent interactions. This alignment with the objective of recommendation tasks ensures that the model captures timely and relevant user preferences, leading to more accurate and personalized recommendations. Importantly, our time-aware regularization approach does not introduce additional embedding encoding. Instead, it dynamically generates graph regularization terms based on the relative time order, making it a lightweight and efficient solution. This design allows the model to handle varying absolute time lengths effectively, showcasing excellent generalization capabilities and requiring minimal computational overhead.

\subsection{Fine-Tuning with Graph \\ Prompt Mechanism}
In this section, we will discuss how we effectively transfer knowledge from a pre-trained Graph Neural Network (GNN) model for fine-tuning with future user-item interactions. To begin the fine-tuning process at the target time $T_n$, which occurs after the pre-training time $T_p$, the intuitive way is to update the model parameters incrementally by simply fine-tuning. That is, we iteratively provide the model with data from the updated time intervals to fine-tune the node representations that were previously updated in the preceding time intervals. Therefore, the initial embeddings for fine-tuning at $T_n$ are derived from the forward pass after the last fine-tuning step.
\begin{equation}
    \mathbf{X}_n^0 = \text{forward}(\mathbf{X}_{n-1}; \mathcal{G}_{n-1}),
\end{equation}
where $\text{forward}(*)$ represents the complete forward pass of the model, utilizing the last fine-tuned embeddings $\mathbf{X}_{n-1}$ and the graph structure $\mathcal{G}_{n-1}$. This method has the advantage of directly capturing users' continuous interest changes within a specific time span. However, the incremental fine-tuning mechanism has two significant drawbacks. First, iteratively updating model parameters based on small-range interactions may lead the model to converge to a local optimum specific to that time period, limiting the potential for continuous fine-tuning on the updated representations in the future. Secondly, persistently updating the parameters of the pre-trained model can result in a significant computational burden.

\subsubsection{\bf Graph-Structural Prompt Mechanism}
In our approach, we address the mentioned issues by leveraging the interaction edges between the pre-training time $T_p$ and the current time $T_n$ as prompt edges. This allows the pretrained model to directly fine-tune on future time periods without the need for iterative updates. Inspired by discrete prompt tuning in large language models~\cite{schick2020exploiting,shin2020autoprompt}, we treat the edges of the graph during a specific time period as discrete prompts that guide the propagation of pretrained embeddings. This captures the representation shift between the pre-training and fine-tuning time points and provides better temporal-aware initial embeddings for fine-tuning. To generate prompt structures, we concatenate the pre-training graph structure with the sampled future edges between the pre-training and current fine-tuning time. This combination enables the model to capture the temporal dynamics and improve the effectiveness of fine-tuning:
\begin{align}
    \mathcal{G}^{\text{prompt}} = \mathcal{G}_p \oplus \sum_{i=1}^{n}\Phi_i\odot\mathcal{G}_{i};\ \Phi_i = \begin{cases}
    1-(i-1)\phi, & \phi > 0 \\
    1+(n-i)\phi, & \phi < 0
    \end{cases},
\end{align}
where "$\oplus$" denotes graph concatenation and "$\odot$" denotes graph sampling. Here, a hyper-parameter $\phi$ is introduced as the sampling decay for prompt structures, where a positive $\phi$ suggests that we include more early structures and less recent ones, and vice versa.
After generating the prompt structures, we proceed with a forward pass using the pretrained embeddings $\mathbf{X}_p$ on the prompt graph to generate embeddings for fine-tuning. To mitigate the overfitting effect and improve generalization for more robust fine-tuning in our \model\ framework, we introduce a random gating~\cite{cai2023ensemble} mechanism that slightly perturbs the pre-trained embeddings.
\begin{align}
\widetilde{\mathbf{X}}_p & = \mathbf{X}_p \odot \text{sigmoid}(\widetilde{\mathbf{W}}\mathbf{X}_p + \widetilde{\mathbf{b}}), \\
    \mathbf{X}^0_n &= \text{forward}(\widetilde{\mathbf{X}}_p; \mathcal{G}^\text{prompt}),
\end{align}
The non-learnable random gating weights, $\widetilde{\mathbf{W}} \in \mathbb{R}^{d\times d}$ and $\widetilde{\mathbf{b}} \in \mathbb{R}^{d}$, are generated from a Gaussian distribution. It's important to note that the relative time encoding also plays a vital role in facilitating the model's ability to sense relative temporal connections during the prompt propagation process. By propagating the embeddings learned from extensive pretraining over a large time period on the prompt edges, which include interactions from subsequent time periods, we achieve two objectives. Firstly, we enable the obtained embeddings to maintain stable user interests. Secondly, we swiftly capture changes in user interests within the subsequent time span. By refraining from directly training the embeddings on the short-term graph, we mitigate the risk of the model parameters becoming trapped in local optima. This approach grants us greater flexibility for subsequent fine-tuning and enables the model to more effectively adapt to users' evolving interests over time.

\subsubsection{\bf Prompt Learning with Adaptive Gating Mechanism}
To address the distribution shift in node representations between the time-aware graph snapshots $\mathcal{G}_{n-1}$ and $\mathcal{G}_{n}$, we introduce a learnable gating mechanism that adaptively transforms the input embeddings $\mathbf{X}_n^0$. This gating mechanism allows for modeling the changes in user/item representations over time, effectively preserving the informative signals necessary for making accurate future recommendations. We employ gradient truncation on $\mathbf{X}_n^0$ to prevent direct optimization of the large-scale pre-trained model. Instead, we fine-tune $\mathbf{X}_n^0$ using newly interaction structual contexts $\mathcal{G}_{n}$ to improve the accuracy of predictions at the target time interval $T_n$:

\begin{align}
    \widetilde{\mathbf{X}}_n^0 &= \mathbf{X}_n^0 \odot \text{sigmoid}(\mathbf{W}_l\mathbf{X}_n^0+\mathbf{b}_l), \\
    \mathbf{X}_n &= \text{forward}(\widetilde{\mathbf{X}}_n^0;\mathcal{G}_{n}),
\end{align}
At this stage, we have derived the user and item representations $\mathbf{X}_n$ for making predictions starting from time $T_n$. In this process, the learnable weights $\mathbf{W}_l$ and $\mathbf{b}_l$, which have the same size as the random gating, are introduced. To estimate the probability of user $u$ interacting with item $i$, we calculate the dot product between the user and item representations $\mathbf{x}_n^u$, $\mathbf{x}_n^i$, denoted as $\hat{y}_{u,i} = {\mathbf{x}_n^u}^\trans \cdot \mathbf{x}_n^i$.

\subsection{Model Learning and Discussion}
\subsubsection{\bf Optimized Objective}
In both the pre-training and fine-tuning stages, we define our training objectives based on optimizing the BPR loss. The BPR loss ensures that the predicted score for an observed interaction is higher than that of its unobserved negative samples. This loss function is commonly used in recommendation systems to model the preference ranking between items for individual users. By optimizing this loss, we aim to improve the model's ability to accurately rank and predict user-item interactions.
\begin{equation}
    \mathcal{L} = - \sum_{(u,i,j) \in D} \log\sigma(\hat{y}_{ui} - \hat{y}_{uj}),
\end{equation}
In our training strategy, we utilize a dataset $D$ that includes negative items $j$ sampled at each training mini-batch. Our approach follows a two-stage process. In the first stage, we pre-train a GNN-based recommender on a large-time-scale graph until convergence. This involves training the model on a comprehensive set of historical data, allowing it to learn long-term patterns and user preferences. In the second stage, we fine-tune the pre-trained model on small-time-scale graph snapshots that include interactions from a more recent time period. This fine-tuning process helps the model adapt and capture short-term changes in user interests and item dynamics.

\subsubsection{\bf{Interplotive Parameter Update}}
To ensure that the model parameters are learned in synchronization with the evolving user and item representations, it is important to update the pre-trained node embeddings over time steps. Inspired by the investigation in~\cite{pareja2020evolvegcn,you2022roland}, we propose an interpolative approach for updating the pre-trained user and item embeddings. Specifically, to estimate the best initial state for training at the next time step $T_n$, we combine the pre-trained embeddings with the embeddings learned within a sliding window $[T_{n-\omega}, T_{n-1}]$ using interpolation. This allows us to leverage both the long-term knowledge captured during pre-training and the recent changes observed within the sliding window, enabling the model to effectively adapt to the evolving dynamics of user-item interactions:
\begin{align}
    \mathbf{X}^{\text{init}}_n = \text{mean}(\mathbf{X}_p,\sum_{i=1}^{\omega}\frac{i\cdot\mathbf{X}_{n-i}}{\sum_{k=1}^{\omega}k})
\end{align}
$\mathbf{X}$ represents the model parameters, which correspond to the user and item embeddings. The left term of the equation calculates a weighted normalization of the weights $[\mathbf{X}_{n-1},...,\mathbf{X}_{n-\omega}]$, where the more recent fine-tuned representations are given less weight. This weighting helps to mitigate the local optima effect, where the model may get stuck in suboptimal solutions based on recent but noisy information. The hyperparameter $\omega$ controls the size of the sliding window, which determines the number of previous time steps considered for fine-tuning. As $\omega$ becomes smaller, the model updates its evolved representations more frequently, allowing it to capture recent changes. However, this can increase the risk of getting trapped in local optima due to the limited historical information considered. On the other hand, if $\omega$ is larger, the model can incorporate longer-term information, but may have reduced sensitivity to recent fine-tuned weights.

%% file: exp.tex
\section{Evaluation}

In this section, we compare our proposed \model\ with state-of-the-art methods to address the following research questions.\vspace{-0.05in}
\begin{itemize}[leftmargin=*]

\item \textbf{RQ1}: Can \model\ outperform other time-aware graph learning models and pre-trained GNNs in dynamic recommendations?

\item \textbf{RQ2}: How does \model\ perform when integrated as a model-agnostic plug-in component with state-of-the-art recommenders?

\item \textbf{RQ3}: Can \model\ perform equally well or even better than the vanilla full-data training paradigm?

\item \textbf{RQ4}: How does the performance of \model\ change under different ablation settings of key components and hyper-parameters?

\item \textbf{RQ5}: How effective is \model\ in tackling the cold-start issue?

\item \textbf{RQ6}: How does the potential scalability of \model\ facilitate efficient model convergence with our prompt learning paradigm?

\item \textbf{RQ7}: Can \model effectively empower real-world recommendation systems when deployed in industrial applications?

\end{itemize}


\vspace{-0.1in}
\subsection{Experimental Settings}
\subsubsection{\bf Datasets}
\begin{table}[]
    \centering
    \caption{Statistics of the experimental datasets.}
    \vspace{-0.15in}
	\resizebox{\linewidth}{!}{
    \begin{tabular}{lcccc}
    \toprule
    Statistics & Taobao & Koubei & Amazon & \\
    \midrule
    \# Users & 117,450 & 119,962 & 131,707 & \\
    \# Items & 86,718 & 101,404 & 107,028 & \\
    \# Interactions & 8,795,404 & 3,986,609 & 876,237 & \\
    \# Density & 8.6e-4 &  3.3e-4 & 6.2e-5 & \\
    \cmidrule(lr){1-5}
    Temporal Segmentation\\
    \cmidrule(lr){1-5}
    \# Pre-training Span & 5 days & 4 weeks & 4 weeks \\
    \# Tuning-Predicting Span & 5 days & 5 weeks & 9 weeks \\
    \# Snapshot Granularity & daily & weekly & weekly & \\
    \hline
    \end{tabular}
    }
    \label{tab:dataset}
    \vspace{-0.2in}
\end{table}


We utilize three public datasets that cover diverse real-world scenarios in dynamic recommendation. The \textbf{Taobao} dataset captures implicit feedback from Taobao.com, a Chinese e-commerce platform, over a period of 10 days. The \textbf{Koubei} dataset, provided for the IJCAI'16 contest, records 9 weeks of user interactions with nearby stores on Koubei in Alipay. The \textbf{Amazon} dataset consists of a 13-week collection of product reviews sourced from Amazon. More details about these datasets can be found in Table~\ref{tab:dataset}.\vspace{-0.05in} 

\subsubsection{\bf Baseline Models}
We include the recent dynamic GNNs and graph prompt approaches as our baselines. Specifically, three most relevant research lines are included for comparison:

\paratitle{Dynamic Recommendation Methods}. 
We include DGCN~\cite{dgcn}, which explores categorizes edges as past and current and designing a GNN to propagate the information from past edges to current ones. However, it does not explicitly consider a dynamic setting with snapshots and focuses on a single graph. \\\vspace{-0.12in}

\paratitle{Graph Prompt Methods}. This line aims to unify the pre-training and downstream tasks using a common template while leveraging prompts for task-specific knowledge retrieval.

\begin{itemize}[leftmargin=*]
\item GraphPrompt~\cite{liu2023graphprompt}. It introduces an approach to pretraining and prompting in the context of graphs. It utilizes a learnable prompt to guide downstream tasks, enabling them to access relevant knowledge from pretrained models using a shared template.

\item GPF~\cite{gpf}. This method introduces prompts within the feature space of the graph, thereby establishing a general approach for tuning prompts in any pre-trained graph neural networks.
\end{itemize}

\paratitle{Dynamic Graph Neural Networks}. 
These networks are tailored to dynamic graphs, updating embeddings with time sensitivity to reflect graph changes. We benchmark our approach against notable models EvolveGCN \cite{pareja2020evolvegcn} and ROLAND \cite{you2022roland}:

\begin{itemize}[leftmargin=*]
    \item EvolveGCN~\cite{pareja2020evolvegcn}. It adapts to graph evolution using recurrent neural networks to modify GCN parameters over time, available as hidden state (-H) or recurrent input (-O) variants.
    
    \item ROLAND~\cite{you2022roland}. A dynamic graph method that employs meta-learning to refresh embeddings for subsequent re-initialization, integrating these with GNN's layer-wise hidden states.
\end{itemize}



\subsubsection{\bf{Integration with GNN Recommenders}}
\model\ is a versatile architecture that seamlessly integrates as a plug-in component with any GNN-based recommender, highlighting its flexibility. In our evaluation, we implement \model\ using the efficient LightGCN~\cite{lightgcn} model. Additionally, we extend \model's applicability by integrating it with self-supervised learning-based recommenders like SGL~\cite{sgl}, MixGCF~\cite{huang2021mixgcf}, and SimGCL~\cite{simgcl}. This integration provides empirical evidence of \model's effectiveness in enhancing dynamic and adaptable recommendations.

\vspace{-0.05in}
\subsubsection{\bf{Evaluation Protocols}} 
In our evaluation, we model real-world dynamics using graph snapshots at varying intervals, such as daily or weekly. A two-step sliding window technique is applied to learn from current data and predict subsequent changes. Under the Pre-train and Fine-tune framework, we pre-train with a substantial dataset fraction, fine-tune, and test on the latter snapshots (refer to Table~\ref{tab:dataset}). Baselines adopt the same approach to maintain consistency. Dynamic GNNs start fine-tuning with weights from the pre-training phase. We average our results over all future temporal snapshots, using standard metrics such as Recall@k and nDCG@k at k=20, in line with existing methodologies \cite{lightgcn, sgl, ncf}.

\begin{table}[t]
\centering
\caption{When compared to various baselines utilizing different backbone architectures, \model\ consistently exhibits strong overall performance across different types of datasets. The script $\ast$ denotes the statistically significant results compared to the second best at $p<0.01$ level.}
\vspace{-0.1in}
\label{tab:overall}
\resizebox{\linewidth}{!}{
\begin{tabular}{lcccccc}
\toprule
\multirow{2}{*}{Method} & \multicolumn{2}{c}{Taobao} & \multicolumn{2}{c}{Koubei} & \multicolumn{2}{c}{Amazon} \\
\cmidrule(lr){2-3}\cmidrule(lr){4-5}\cmidrule(lr){6-7}
~ & Recall & nDCG & Recall & nDCG & Recall & nDCG \\
\midrule
DGCN & 0.0229 & 0.0228 & \underline{0.0353} & \underline{0.0255} & 0.0158 & 0.0084 \\
\midrule
\textbf{LightGCN+} \\
\cmidrule(lr){1-7}
GraphPrompt  & 0.0199 & 0.0195 & 0.0342 & 0.0249 & 0.0154 & 0.0075 \\
GPF & 0.0223 & 0.0220 & 0.0348 & {0.0251} & \underline{0.0174} & \underline{0.0088} \\
EvolveGCN-H & 0.0224 & 0.0221 & 0.0315 & 0.0231 & 0.0138 & 0.0066 \\
EvolveGCN-O & \underline{0.0236} & \underline{0.0232} & 0.0334 & 0.0242 & 0.0157 & 0.0084 \\
ROLAND & 0.0226 & 0.0226 & 0.0301 & 0.0223 & 0.0150 & 0.0069 \\
\textbf{\model}\ & \textbf{0.0251}$^\ast$ & \textbf{0.0245}$^\ast$ & \textbf{0.0362}$^\ast$ & \textbf{0.0265}$^\ast$ & \textbf{0.0191}$^\ast$ & \textbf{0.0094}$^\ast$ \\
\midrule
\textbf{SGL+} \\
\cmidrule(lr){1-7}
GraphPrompt & 0.0223 & 0.0220 & 0.0355 & 0.0261 & 0.0161 & 0.0079 \\
GPF & 0.0229 & 0.0226 & 0.0363  & 0.0266 & \underline{0.0187} & \underline{0.0096} \\
EvolveGCN-H & 0.0235 & 0.0232 & 0.0358 & 0.0263 & 0.0137 & 0.0066 \\
EvolveGCN-O & \underline{0.0242} & \underline{0.0238} & \underline{0.0365} & \underline{0.0268} & 0.0173 & 0.0090 \\
ROLAND & 0.0222 & 0.0222 & 0.0340 & 0.0251 &  0.0161 & 0.0078 \\
\textbf{\model}\ & \textbf{0.0268}$^\ast$ & \textbf{0.0264}$^\ast$ & \textbf{0.0371}$^\ast$  & \textbf{0.0277}$^\ast$ & \textbf{0.0221}$^\ast$ & \textbf{0.0114}$^\ast$ \\
\midrule
\textbf{MixGCF+} \\
\cmidrule(lr){1-7}
GraphPrompt & 0.0248 & 0.0245 & 0.0377 & 0.0276 & 0.0180 & 0.0089 \\
GPF & 0.0251 & 0.0247 & \underline{0.0380}  & \underline{0.0278} &  \underline{0.0182} & \underline{0.0092} \\
EvolveGCN-H & 0.0240 & 0.0237 & 0.0354 & 0.0262 & 0.0129 & 0.0061 \\
EvolveGCN-O & \underline{0.0271} & \underline{0.0267} & 0.0375 & 0.0276 & 0.0171  & 0.0085 \\
ROLAND & 0.0232 & 0.0230 & 0.0349 & 0.0260 &  0.0152 & 0.0072 \\
\textbf{\model}\ & \textbf{0.0280}$^\ast$ & \textbf{0.0273}$^\ast$ & \textbf{0.0393}$^\ast$ & \textbf{0.0291}$^\ast$ & \textbf{0.0216}$^\ast$ & \textbf{0.0109}$^\ast$ \\
\midrule
\textbf{SimGCL+} \\
\cmidrule(lr){1-7}
GraphPrompt & 0.0239 & 0.0224 & 0.0348 & 0.0258 &  0.0139 & 0.0069 \\
GPF & 0.0237 & 0.0220 & 0.0357 & 0.0264 &  \underline{0.0182} & \underline{0.0094} \\
EvolveGCN-H & 0.0241 & 0.0238 & \underline{0.0356} & \underline{0.0265} & 0.0134 & 0.0067 \\
EvolveGCN-O & \underline{0.0241} & \underline{0.0238} & 0.0351 & 0.0258 & 0.0168  & 0.0088 \\
ROLAND & 0.0228 & 0.0228 & 0.0333 &  0.0246 & 0.0151 & 0.0075 \\
\textbf{\model}\ & \textbf{0.0280}$^\ast$ & \textbf{0.0276}$^\ast$ & \textbf{0.0368}$^\ast$ & \textbf{0.0276}$^\ast$ & \textbf{0.0205}$^\ast$  & \textbf{0.0108}$^\ast$ \\
\bottomrule
\end{tabular}
}
\vspace{-0.2in}
\end{table}

\vspace{-0.1in}
\subsection{Performance Comparison (RQ1--RQ3)}

\subsubsection{\bf{Comparison with Baselines}}
We showcase the performance of \model\ alongside other methods, such as graph prompt techniques and dynamic GNNs, using LightGCN as the baseline, detailed in Table~\ref{tab:overall}. Key observations from the analysis include:

\begin{itemize}[leftmargin=*]
    \item Our \model\ surpasses both graph prompt and dynamic graph learning methods, highlighting the efficacy of our pre-training and prompt learning strategy. This performance can be credited to: 1) Our temporal prompt mechanism that adeptly captures the evolving user-item interactions during pre-training and fine-tuning, and 2) Our graph prompt design that ensures seamless knowledge transfer from the pre-trained model and mitigates distribution shifts across temporal snapshots. \\\vspace{-0.12in}


    \item The varying performance of baseline methods underscores the complexity of dynamic recommendations. EvolveGCN excels with the Taobao dataset but not universally, possibly due to overfitting short-term user behaviors with its complex architecture. In contrast, \model\ employs streamlined prompt mechanisms that adeptly seize long-term user interests and assimilate fresh behavioral data, offering a more consistent and effective approach. \\\vspace{-0.12in}


    \item Despite being regarded as the meticulously-designed dynamic GNN, ROLAND does not demonstrate superior performance. This limitation may be attributed to its intricate model parameter update schemes, which introduce larger perturbations to embeddings. Consequently, the representation learning for users and items is disrupted, rendering it less effective in capturing the time-evolving user preferences in recommendation tasks.

\end{itemize}

\vspace{-0.05in}
\subsubsection{\bf{Integration with SOTA Methods}}
We also evaluate \model's versatility with various base recommenders—MixGCF, SGL, and SimGCL—reimplementing all methods under uniform evaluation criteria. The averaged results across time slots can be found in Table~\ref{tab:overall}. Our summarized observations are as follows:

\begin{itemize}[leftmargin=*]

    \item \model\ consistently excels alongside advanced recommenders, showcasing our method's flexibility and capability to elevate performance across scenarios. Baselines, however, show varied performance depending on the base recommender and dataset used, with no single method dominating across all scenarios. While EvolveGCN-O often ranks second on Taobao, GPF stands out on Amazon regardless of the base recommender. Such variability in response to dataset specifics and underlying models prevents baselines from achieving consistent, substantial outcomes.

    \item Superior base model representations typically yield better results. \model\ sees a 4.5\% boost when comparing SGL to SimGCL on Taobao, suggesting it capitalizes well on the base model's improved representations in dynamic contexts. Conversely, methods like EvolveGCN-O fail to show gains with better base models, hinting at potential generalization weaknesses in their design.

\end{itemize}
\vspace{-0.05in}
\subsubsection{\bf Comparison with Full-Data Training} We place the results and discussion in Appendix~\ref{app:full} due to space limitation.
\vspace{-0.1in}
\subsection{Ablation Study (RQ4)}
\subsubsection{\bf{Key Components in Fine-Tuning}}
\begin{figure}[t]
\centering
\subfigure[Taobao]{
\label{fig:ab:taobao}
\includegraphics[width=0.31\linewidth]{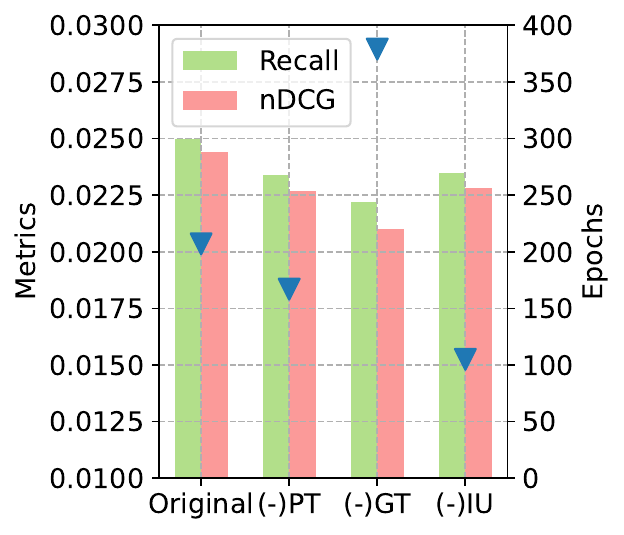}}
\subfigure[Koubei]{
\label{fig:ab:koubei}
\includegraphics[width=0.31\linewidth]{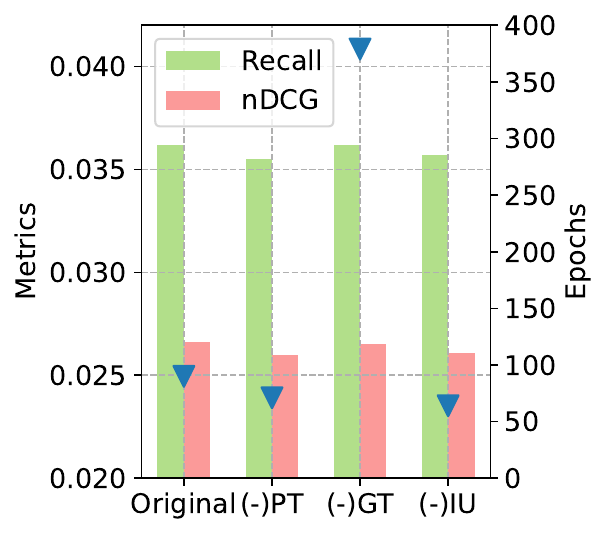}}
\subfigure[Amazon]{
\label{fig:ab:amazon}
\includegraphics[width=0.31\linewidth]{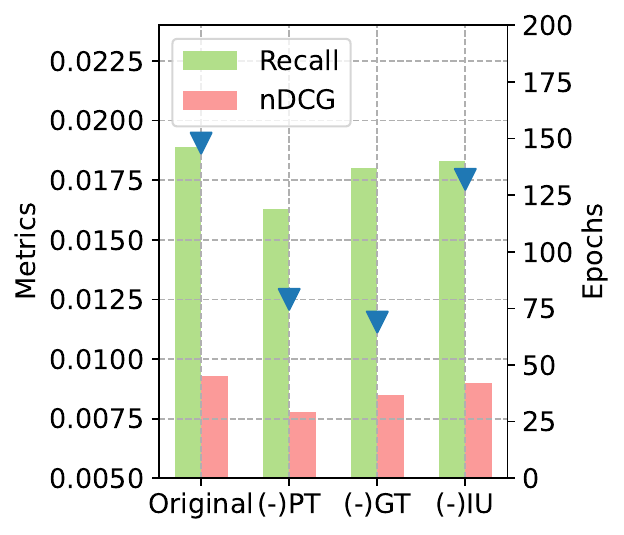}}
\vspace{-0.2in}
\caption{\textbf{Key component ablation study for fine-tuning stage. Y-axis denotes performance metrics on the left and epochs (displayed as $\triangledown$) for convergence on the right.}}
\label{fig:ab}
\vspace{-0.2in}
\end{figure}

We perform a detailed ablation study to assess the impact of \model's crucial elements during pre-training and fine-tuning. For comparative analysis against the original design, we introduce three \model\ variants, each omitting a specific key component:
\begin{itemize}[leftmargin=*]
    \item (-)PT: This variant removes the \textbf{P}rompt \textbf{T}uning component that leverages prompt edges from past interactions, opting to fine-tune the model directly with new interactions.
    \item (-)GT: Adaptive \textbf{G}a\textbf{T}ing mechanism, which facilitates dynamic knowledge transformation in the fine-tuning phase, is omitted.
    \item (-)IU: The \textbf{I}nterplotive \textbf{U}pdate module is excluded, maintaining static pre-trained weights during fine-tuning.
\end{itemize}
From the data in Figure~\ref{fig:ab}, we discern that: 1) Each of the three integral elements of \model\ plays a vital role. Omitting any one of them not only diminishes recommendation precision but may also prolong the time to convergence, underscoring their collective efficacy. 2) The structural prompt and interpolative update features are instrumental in boosting accuracy and avoiding local optima during dynamic learning. Their absence can lead to quicker convergence but at the cost of significantly reduced accuracy. 3) The adaptive gating feature is key to expedited convergence and enhanced accuracy through improved gradient flow. Without it, we observe notably extended convergence times and diminished accuracy, highlighting its role in bridging the distribution shifts encountered during fine-tuning across snapshots.

\vspace{-0.05in}
\subsubsection{\bf{Effect of Pre-trained Model}}
\begin{figure}[t]
\centering
\subfigure[Taobao]{
\label{fig:pre_ab:taobao}
\includegraphics[width=0.48\linewidth]{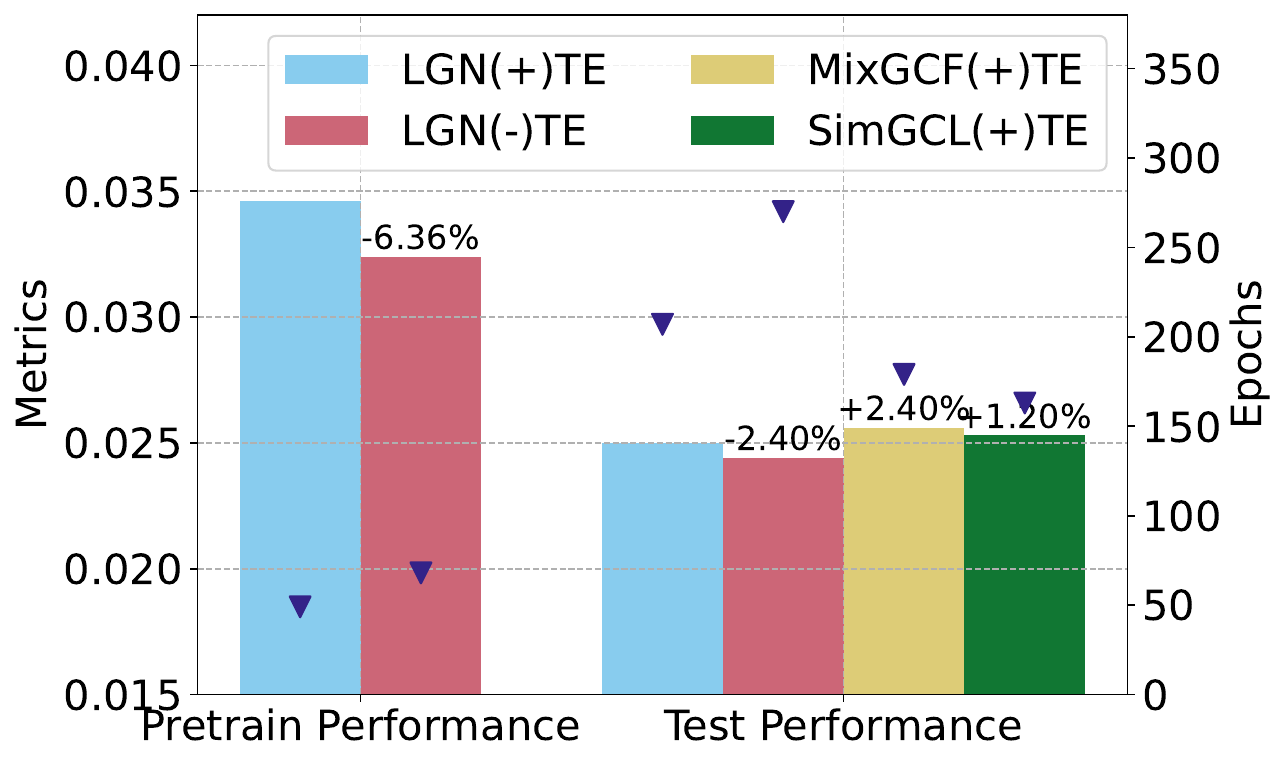}}
\subfigure[Koubei]{
\label{fig:pre_ab:koubei}
\includegraphics[width=0.48\linewidth]{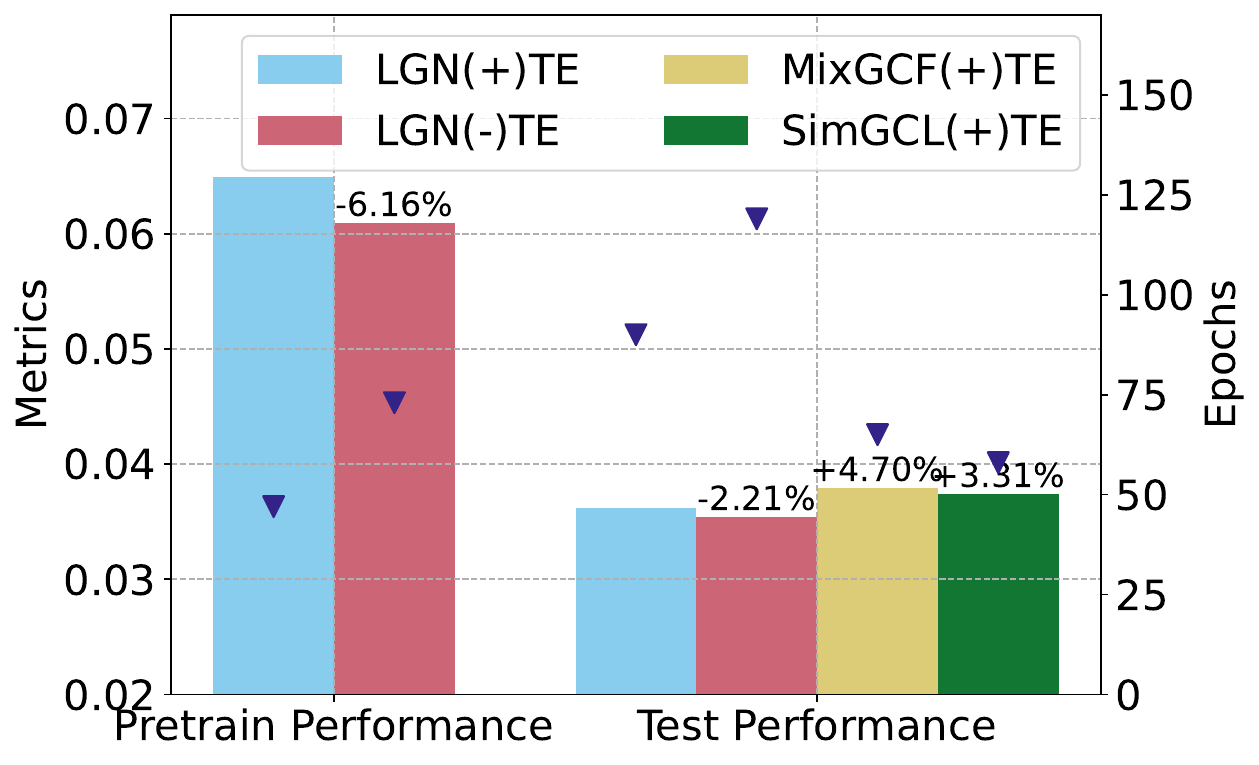}}
\vspace{-0.2in}
\caption{\textbf{Ablation study for pretrained models. }}
\label{fig:ab_pre}
\vspace{-0.15in}
\end{figure}

We assess how different pretraining model architectures influence fine-tuning outcomes by examining four variants, focusing on the use of relative time encoding and the intrinsic representational strength of the models. The baseline model "LGN(+)TE" incorporates time encoding (TE) with LightGCN. The "LGN(-)TE" variant excludes TE during pretraining. The other two, "MixGCF(+)TE" and "SimGCL(+)TE," employ more robust models for pretraining while maintaining LightGCN for fine-tuning. Performance contrasts between pretraining and fine-tuning for these variants are depicted in Figure~\ref{fig:ab_pre}.
\begin{itemize}[leftmargin=*]


\item The temporal prompt mechanism enhances both the speed of convergence and the accuracy of predictions during pretraining and fine-tuning. It directs the model to capitalize on crucial temporal cues in the message-passing process.

\item The use of more robust pretrained models correlates with improved performance, echoing trends seen in pre-trained language models like BERT~\cite{devlin2018bert} and vision models like ViT~\cite{dosovitskiy2020vit}. Our framework's scalability and adaptability are evident, as it allows the use of powerful pretrained models to excel in recommendation. This underscores the strategy of refining large pretrained models with more streamlined models downstream for enhanced results.

\end{itemize}

\vspace{-0.1in}
\subsection{Learning Impact Analysis (RQ5 \& RQ6)}
\subsubsection{\bf{Fine-tuned v.s. Untuned (Cold-start) Nodes}}
\begin{figure}[t]
\centering
\subfigure[Tuned Users]{
\label{fig:group:tuned}
\includegraphics[width=0.48\linewidth]{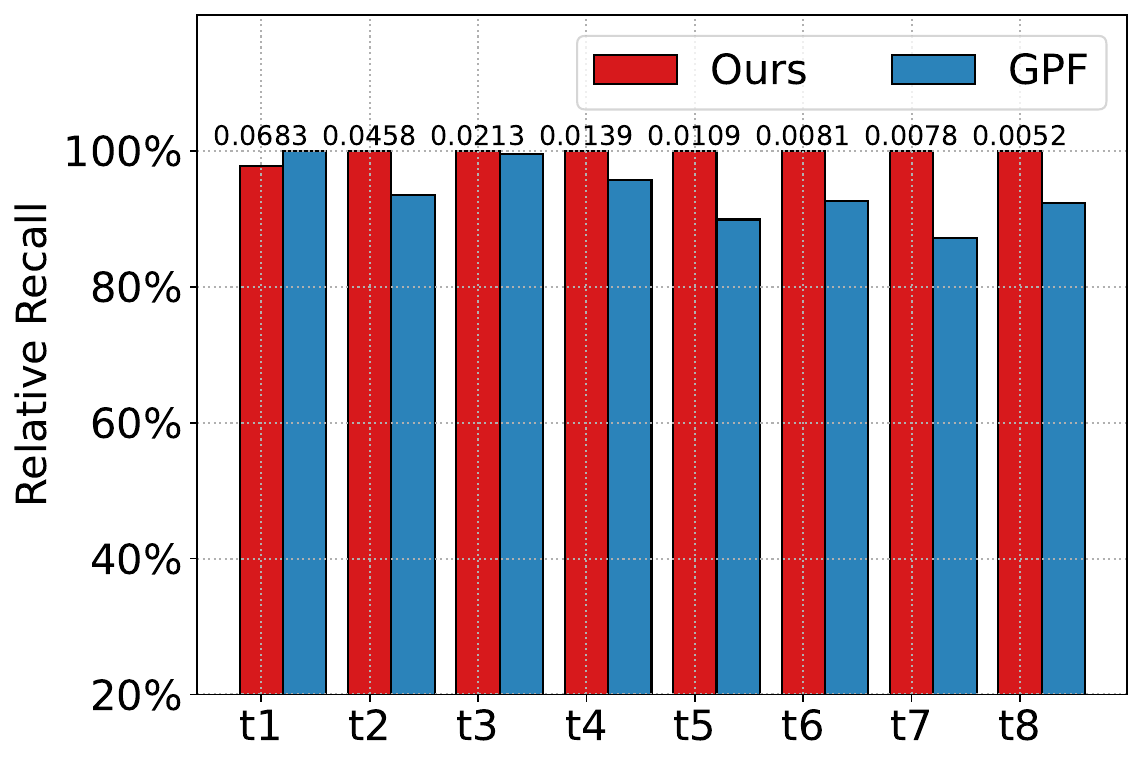}}
\subfigure[Untuned Users]{
\label{fig:group:untuned}
\includegraphics[width=0.48\linewidth]{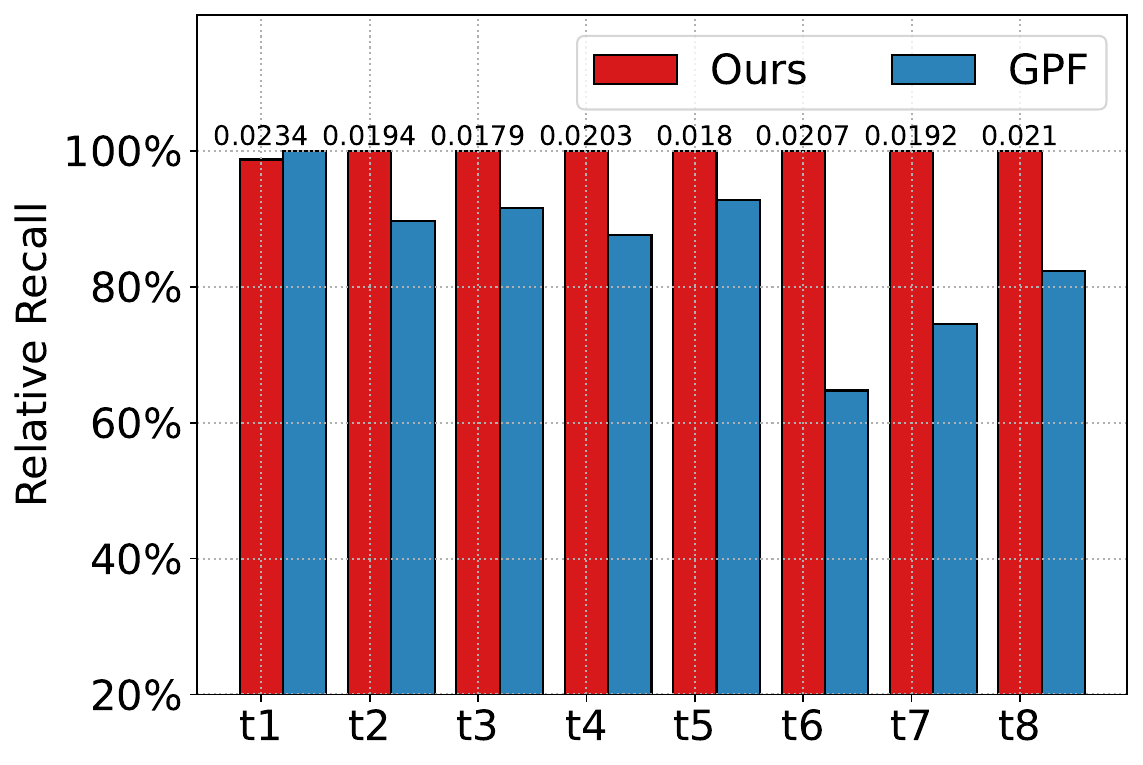}}
\vspace{-0.2in}
\caption{Evaluation performance for tuned and untuned users on Amazon compared with the best baseline, GPF.}
\label{fig:tuned}
\vspace{-0.1in}
\end{figure}

This section delves into the advantages of our fine-tuning approach on node representation learning for recommendations. We divide users into two categories according to whether they are subject to fine-tuning at each time step. Users who do not undergo fine-tuning are considered as cold-start users for that period. Using the Amazon dataset, we assess these two groups independently at every time step and present the findings in Figure~\ref{fig:tuned}.
\begin{itemize}[leftmargin=*]

\item \model\ effectively enhances representations for both tuned and cold-start users, consistently outperforming in most scenarios. This success is largely due to the structural prompt design, which leverages past interactions to enrich the representations.

\item \model\ showcases enhanced performance over the baseline with sustained long-term improvement. After an initial lag in the first snapshot, \model\ overtakes the top baseline from $T_2$ through $T_8$, underscoring its effectiveness at navigating local optima for superior long-term results in dynamic settings.

\end{itemize}

\vspace{-0.05in}
\subsubsection{\bf{Efficiency in Learning}}
\begin{figure}[t]
\centering
\subfigure[Taobao]{
\label{fig:curve:taobao}
\includegraphics[width=0.48\linewidth]{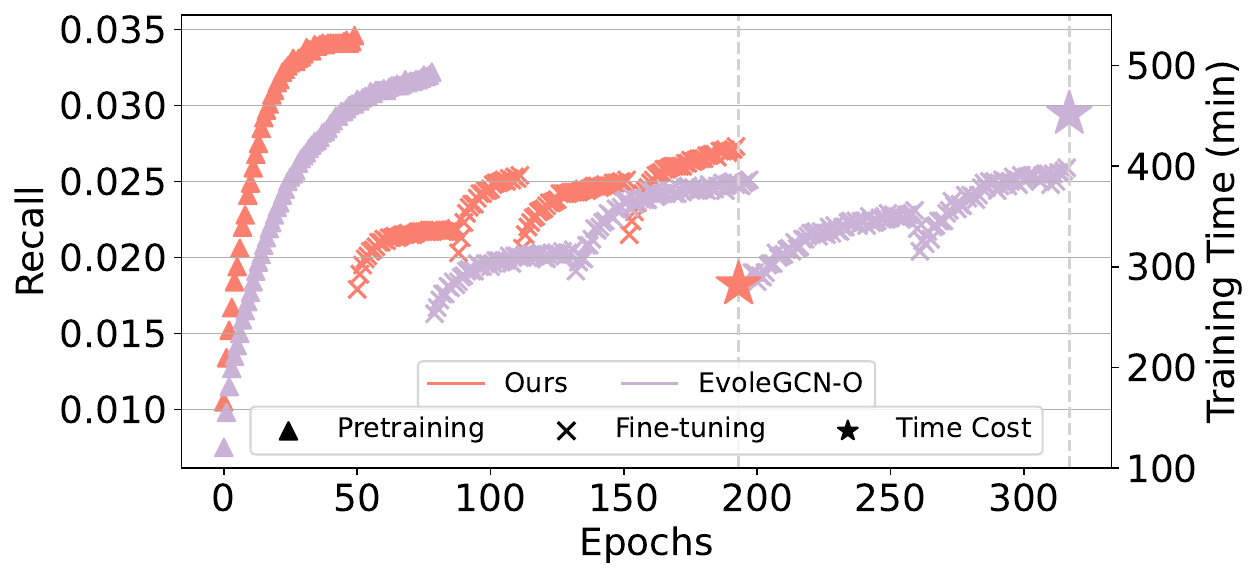}}
\subfigure[Koubei]{
\label{fig:curve:koubei}
\includegraphics[width=0.48\linewidth]{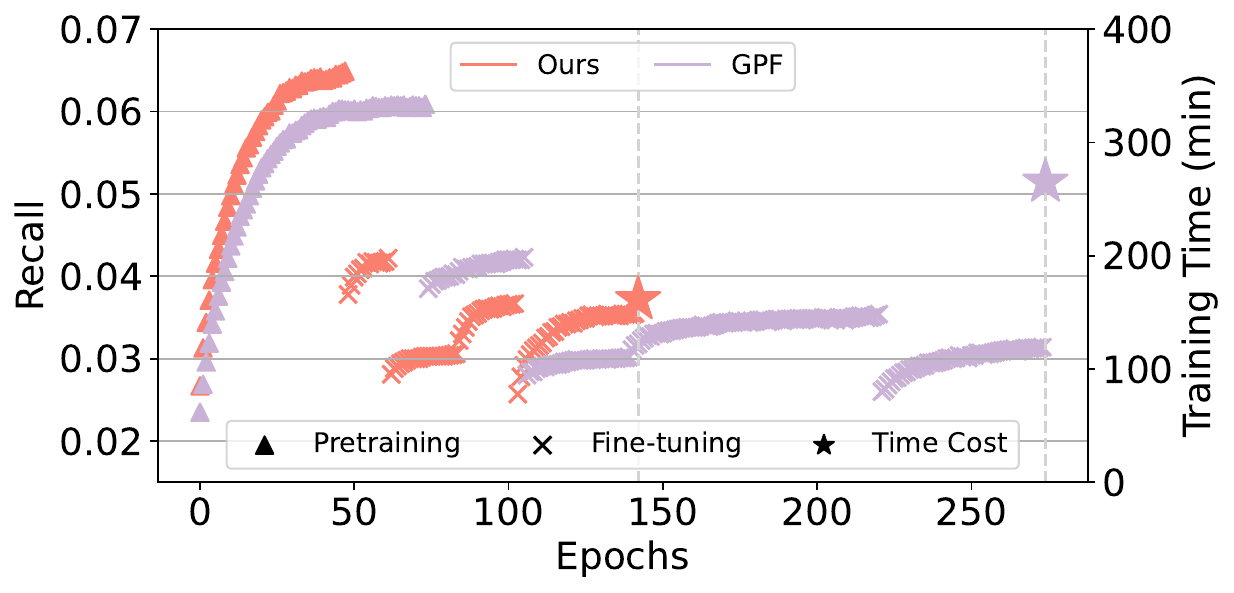}}
\vspace{-0.2in}
\caption{The training curves for \model and baselines on the Taobao and Koubei are presented, with scatters marking performance over various stages (pre-training and fine-tuning) against epochs. A star marker denotes the final convergence, and the right y-axis shows the total time consumed.}
\label{fig:curve}
\vspace{-0.1in}
\end{figure}

This section examines the learning efficiency of our \model. As a parameter-efficient approach, \model minimizes learnable weights during both pretraining and fine-tuning, unlike incremental training methods, thereby reducing training time and computational costs. Our comparisons of training curves, shown in Figure~\ref{fig:curve}, reveal that \model not only outperforms EvolveGCN-O and GPF on the Taobao and Koubei datasets but also achieves this with fewer epochs. For example, \model reaches convergence after just four fine-tuning stages, taking roughly half the epochs and time compared to the baselines, underscoring the efficiency improvements \model provides.

\subsection{Online Deployment (RQ7)}
\model\ is implemented on a major online content platform (name omitted for anonymity), serving millions of users, to assess its effectiveness in personalizing content recommendations. Integrating \model into the core CTR prediction system, it utilizes unsupervised DGI-trained user embeddings as pretraining weights. Historical item-to-user interactions shape prompt edges, and these embeddings are fine-tuned with a 1-layer GNN to inform item representations for the primary model, with updates occurring every ten minutes. During an online A/B test, \model and the current online model each engage approximately 2 million users. Performance is evaluated over a 5-day period using four metrics related to CTR and click count, as presented in Table~\ref{tab:online}. Results show that \model significantly enhances the real-world recommender system by effectively modeling evolving user and item representations and leveraging deep user interests through pretraining and fine-tuning, all while being easy and cost-effective to deploy.

\begin{table}[t]
\centering
\caption{Online A/B test results spanning 5 days. HPC: highly-personalized content. CC: click count. VCC: video click count.}
\vspace{-0.15in}
\resizebox{\linewidth}{!}{
\begin{tabular}{lcccc}
\toprule
\textbf{Model} & \textbf{CTR} & \textbf{HPC CTR}& \textbf{Avg. CC} & \textbf{Avg. VCC} \\
\midrule
Online Model & 10.61\% & 13.42\% & 0.6716 & 0.0188 \\
\model & 10.78\% & 13.89\% & 0.6831 & 0.0194 \\
\# Improve & 1.53\%$\pm$ 0.68\% & 3.45\%$\pm$ 0.64\% & 1.71\%$\pm$ 0.84\% & 3.28\%$\pm$ 1.76\% \\
\bottomrule
\end{tabular}}
\vspace{-0.15in}
\label{tab:online}
\end{table}

%% file: relate.tex
\section{Related Works}
\paratitle{GNNs for Recommendation.}
GNNs have become key in recommendation systems for capturing multi-hop collaborative signals~\cite{gao2023survey}. Models like NGCF~\cite{ngcf} and PinSage~\cite{pinsage} recursively enhance user and item embeddings through message-passing. GCCF~\cite{gccf} adds a residual structure, whereas LightGCN~\cite{lightgcn} streamlines by discarding non-linearities. Further innovations include hypergraph learning~\cite{li2022enhancing,yang2022multi} and intent disentanglement~\cite{wu2022disentangled,ren2023disentangled} to parse complex collaborative patterns. To tackle sparsity in graph recommenders, self-supervised techniques such as contrastive~\cite{li2023self, shuai2022review} and generative approaches~\cite{xia2023automated} have been introduced. \\\vspace{-0.12in}

\paratitle{Pre-training and Fine-tuning on GNNs.}
Drawing from NLP's pre-training and fine-tuning success, researchers have applied similar concepts to GNNs. Techniques like contrastive learning and infomax-based pre-training have been designed for advanced representation learning \cite{hu2019strategies, graphcl, dgi, graphinfo}. Innovations in pre-training, such as link prediction, feature generation, and prompt-based fine-tuning, have emerged \cite{gpt-gnn, hou2022graphmae, gpf, liu2023graphprompt, allinone, sun2022gppt}. Yet, these methods haven't fully overcome the intricacies of dynamic graph learning, often resulting in less-than-ideal performance in tasks with temporal elements. Our \model stands out for its proficiency in handling dynamic graphs within the pre-training and fine-tuning.
In the realm of recommendation tasks, \cite{hao2021pre, liu2023graph} suggest pre-training models tailored for user and item modeling. \cite{hao2021pre} pre-trains a GNN to mimic cold-start conditions, while \cite{liu2023graph} develops a pre-training strategy using side information. Nevertheless, these approaches concentrate on static recommendation settings and fail to account for the dynamic evolution of user preferences, thus limiting their effectiveness as time-sensitive recommenders. \\\vspace{-0.12in}

\paratitle{Dynamic Graph Learning.}
Research on learning with temporally evolving graphs is gaining traction. Techniques such as EvolveGCN \cite{pareja2020evolvegcn}, Dyngraph2vec \cite{goyal2020dyngraph2vec}, DGNN \cite{dgnn}, ROLAND \cite{you2022roland}, and WinGNN \cite{zhu2023wingnn} leverage RNNs and recurrent layers to grasp graph changes. Yet, they miss a dedicated pre-training and fine-tuning strategy specific to dynamic graphs and can introduce noise into user and item representations. DGCN \cite{dgcn} does consider dynamics in graph recommendation learning, but it falls short of handling graphs with evolving snapshots, limiting its assessment to a static graph. \\\vspace{-0.12in}

\paratitle{Sequential Recommendation.}
Sequential recommendation, a field attuned to time-sensitive recommendation scenarios, showcases works like attention-based approaches: SASRec \cite{sasrec} and BERT4Rec~\cite{sun2019bert4rec}, GNN-based methods: SURGE~\cite{chang2021sequential} and Retagnn~\cite{hsu2021retagnn}, and SSL models such as S3-rec~\cite{zhou2020s3}, ICL~\cite{chen2022intent}, and DCRec~\cite{yang2023debiased}. Our approach, while similarly embracing temporal dynamics, diverges from these next-item predictors. Sequential models, with their auto-regressive encoders, face limitations in next-item prediction and differ from graph-based methods that excel in top-K item retrieval. Furthermore, they don't incorporate explicit time intervals like days or weeks, but rather rely on fixed historical sequences.

%% file: appendix.tex
\appendix \section{Appendix}
\label{sec:appendix}


\subsection{Hyper-parameter Sensitivity}
\begin{figure}[H]
\centering
\subfigure[Time Interval $\tau$]{
\label{fig:hp:time}
\includegraphics[width=0.31\linewidth]{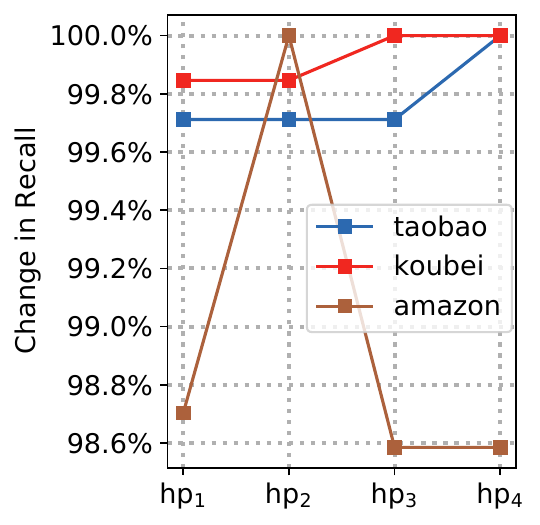}}
\subfigure[Updating Window $\omega$]{
\label{fig:hp:update}
\includegraphics[width=0.31\linewidth]{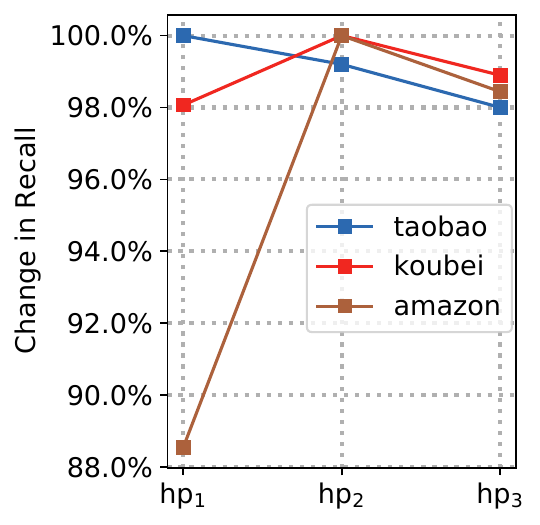}}
\subfigure[Sampling Decay $\phi$]{
\label{fig:hp:sample}
\includegraphics[width=0.31\linewidth]{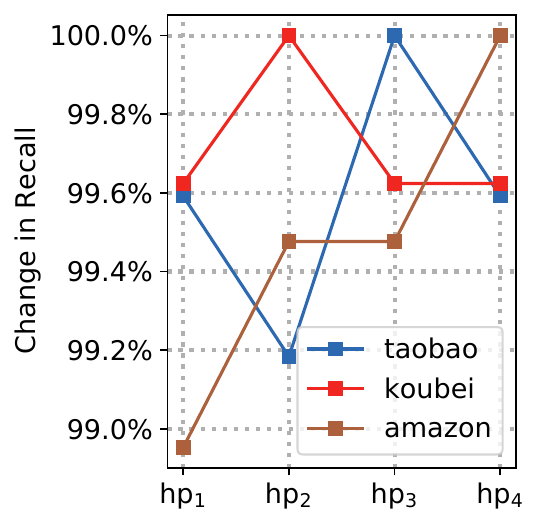}}
\vspace{-0.1in}
\caption{Performance change w.r.t. key hyperparameters.}
\label{fig:hp}
\end{figure}

Our research focuses on investigating the response of \model\ to hyper-parameter adjustments, using three distinct datasets. We conduct a meticulous analysis of the primary hyperparameters that play a crucial role in the operation of \model. These hyperparameters include the temporal prompt interval $\tau$, the updating window $\omega$, which determines the intervals for model updates, and the sampling decay $\phi$, which influences the emphasis on past interactions. To ensure a comprehensive evaluation, we carefully select ranges of values for each hyperparameter that are tailored to the unique characteristics of each experimental dataset utilized.
\begin{itemize}[leftmargin=*]
    \item $\tau$: Taobao-$[0.5, 1, 4, 12]$; Koubei and Amazon-$[24, 48, 72, 96]$
    \item $\omega$: Taobao and Koubei-$[1, 2, 3]$; Amazon-$[2, 4, 6]$
    \item $\phi$: $[0.05, 0.1, -0.05, -0.1]$
\end{itemize}
Our analysis of hyper-parameter sensitivity in \model is presented in Figure~\ref{fig:hp}, from which we derive several observations: \\\vspace{-0.12in}

\noindent 1) \model exhibits a greater degree of sensitivity to hyperparameter changes when applied to the Amazon data in comparison to those of Taobao and Koubei. This heightened sensitivity on Amazon is likely due to its considerably longer fine-tuning and prediction period, which spans an entire 9 weeks. This duration is substantial when contrasted with the other datasets and suggests a correlation between the span of fine-tuning and the impact of hyperparameters on the model's long-term performance. For enhanced long-term accuracy, we need to be careful when picking these settings. \\\vspace{-0.12in}

\noindent 2) When examining the details, it becomes evident that \model is notably more affected by the updating window $\omega$ compared to other hyperparameters in its configuration. The size of the updating window requires careful consideration, as it is crucial to strike a delicate balance. If the window is too narrow, the model may fail to incorporate important updates that contribute to the evolving representations of users and items. On the other hand, if the window is too wide, the model may become burdened with irrelevant noise, thus obscuring the current representation learning process. \\\vspace{-0.12in}

\noindent 3) Moreover, our findings reveal a reassuring aspect of \model's design: its resilience. Minor modifications in the hyper-parameter settings do not precipitate substantial declines in the model's performance. We emphasize the significance of customizing hyperparameter choices to align with the intrinsic characteristics and unique aspects of the dataset being utilized.

\begin{figure}[h]
\centering
\subfigure[Taobao \& Koubei]{
\label{fig:full:1}
\includegraphics[width=0.9\linewidth]{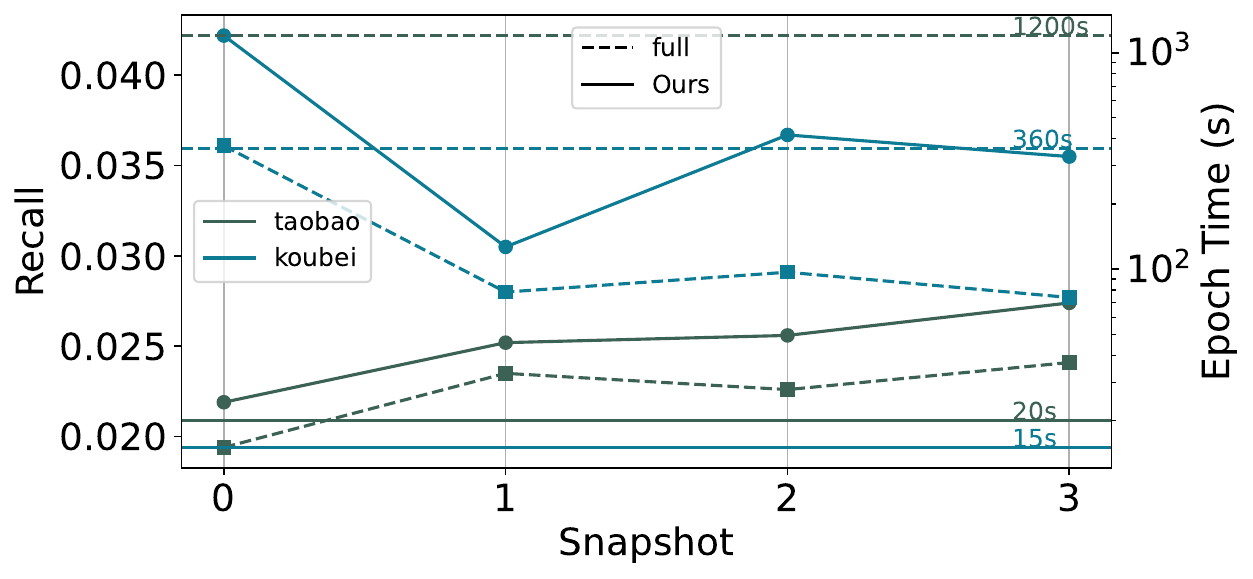}}
\subfigure[Amazon]{
\label{fig:full:2}
\includegraphics[width=0.9\linewidth]{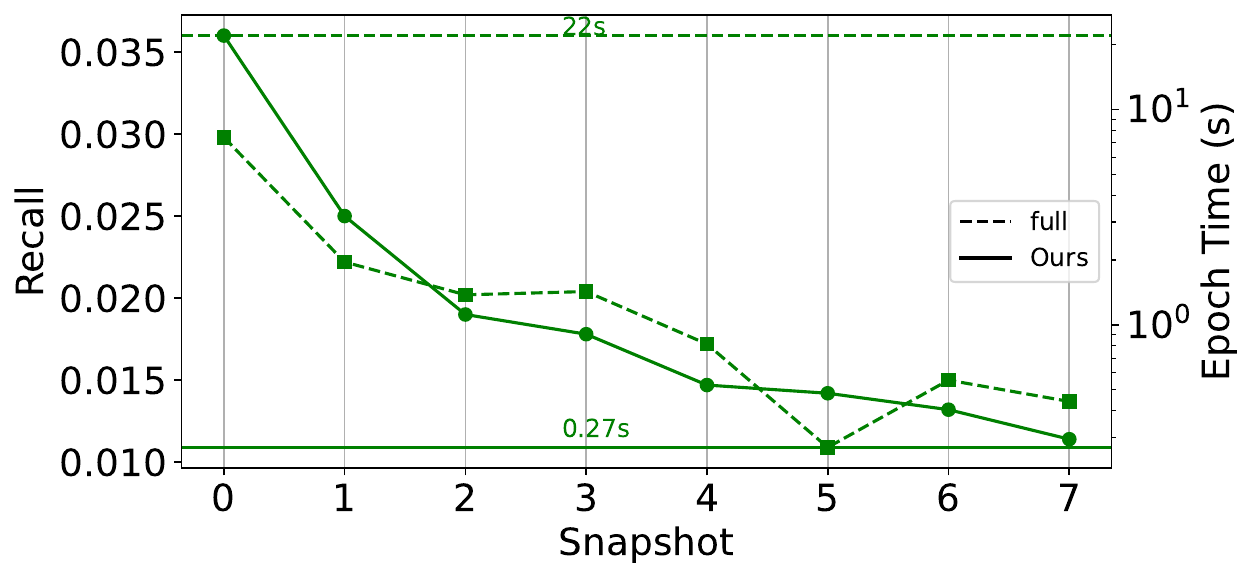}}
\vspace{-0.2in}
\caption{\model\ v.s. full-data training using LightGCN. The results with the average training time for a single epoch are being depicted by a horizontal line on the right Y-axis.
}
\label{fig:full}
\vspace{-0.2in}
\end{figure}

\subsection{Comparsion with Full-Data Training}
\label{app:full}
In this section, our primary objective is to conduct a comprehensive performance comparison between our \model\ and the traditional full-data training approach (FULL). Our aim is to determine if \model\ can match or exceed the performance of FULL. To achieve this, we conduct a meticulous comparison using three distinct datasets, with a keen focus on Recall and average epoch time. The results of our analysis are visually represented in Figure~\ref{fig:full}. Our findings can be summarized into two key observations: \\\vspace{-0.12in}

\noindent 1) In our study, \model\ stands out on both the Taobao and Koubei datasets, surpassing FULL's performance at every testing point. This underscores \model's robust capability in processing time-sensitive data to yield more precise recommendations. With the Amazon dataset, the two approaches are closely matched, but \model\ shows a distinct advantage in the initial phases. This suggests that \model's dynamic method can be immediately beneficial, improving performance from the start without losing ground to FULL as the evaluation progresses over time. \\\vspace{-0.12in}

\noindent 2) Furthermore, a significant aspect of \model's appeal lies in its exceptional efficiency. When compared to the FULL approach, \model\ cuts down the average training time dramatically, achieving astonishing efficiency gains. Specifically, we observe an impressive 60-fold increase in speed on the Taobao dataset, a 24-fold increase on the Koubei dataset, and an unparalleled 81-fold increase on the Amazon dataset. These improvements in training efficiency can be attributed to the innovative architecture of \model, which seamlessly integrates pre-training and prompt learning techniques. This strategic integration significantly accelerates the learning process, enabling \model\ to rapidly assimilate and adeptly apply new data to generate accurate and timely recommendations.